# Enhancing generalization in high-energy physics using white-box adversarial attacks

Franck Rothen,[*] Samuel Klein,[†] Matthew Leigh,[‡] and Tobias Golling[§]

*Department of Particle Physics, Université de Genève, Geneva 1205, Switzerland*

Machine learning is becoming increasingly popular in the context of particle physics. Supervised learning, which uses labeled Monte Carlo (MC) simulations, remains one of the most widely used methods for discriminating signals beyond the Standard Model. However, this paper suggests that supervised models may depend excessively on artifacts and approximations from Monte Carlo simulations, potentially limiting their ability to generalize well to real data. This study aims to enhance the generalization properties of supervised models by reducing the sharpness of local minima. It reviews the application of four distinct white-box adversarial attacks in the context of classifying Higgs boson decay signals. The attacks are divided into weight-space attacks and feature-space attacks. To study and quantify the sharpness of different local minima, this paper presents two analysis methods: gradient ascent and reduced Hessian eigenvalue analysis. The results show that white-box adversarial attacks significantly improve generalization performance, albeit with increased computational complexity.

## I. INTRODUCTION

High-energy physics (HEP) analyses often need to significantly reduce the contribution from different backgrounds, necessitating discriminating signal from background [1,2]. Supervised machine learning, in particular, is gaining popularity as a general tagger for signal events [3–17], capable of utilizing the full range of features available in the data. Complex patterns do not require predefinitions, as the model can learn them directly from the data. For instance, rare Beyond Standard Model (BSM) and background quantum chromodynamics (QCD) processes can be generated using Monte Carlo simulation. These labeled datasets can then be used to train a neural network, which is subsequently applied to real-world data [18]. This study shows that supervised models can be sensitive to specific parameters in Monte Carlo simulation, potentially leading to poor generalization to real data. Therefore, this paper focuses on investigating the generalization capabilities of neural networks in the context of HEP and exploring techniques to improve said performance.

A similar adversarial approach to this problem was explored in Ref. [19], where the fast gradient sign method (FGSM) was used to increase robustness. A key difference in the present study is the use of two different MC simulators to cross-validate generalization performance instead of an adversarial test set. This study also provides a quantitative analysis of sharpness reduction and compares additional adversarial training strategies beyond the FGSM.

This paper is structured as follows: The formalism behind generalization is briefly introduced in Sec. II. The concept of loss sharpness and how adversarial attacks can reduce it is presented in Sec. III, where the implemented adversarial training strategies are also presented. In order to quantify and evaluate the effectiveness of the adversarial techniques in reducing loss sharpness, the paper introduces two evaluation methods in Sec. IV. The experimental setup is detailed in Sec. V. All results are presented in Sec. VI. Finally, the paper concludes in Sec. VII.

## II. GENERALIZATION PERFORMANCES AND CORRELATION WITH SHARPNESS

Following the formalism for generalization from Ref. [20], consider a training dataset composed of $n_e$ subdatasets $e$:

$$D \coloneqq \{(X^e, Y^e) | \ \forall \ e \in \mathcal{E}_{\text{Tr}}\}, \tag{1}$$

where $\mathcal{E}_{\text{Tr}}$ is the set containing all considered training environments. In this case, these environments are generated using MC. The different environments $e$ can be seen as different physical processes, different detector responses, or different approximations. Each environment is independently distributed according to a respective probability density distribution $P(X^e, Y^e)$. The goal of the model is to generalize to new environments $\mathcal{E}_{\text{all}} \supset \mathcal{E}_{\text{Tr}}$ that were not present during training. Most importantly, generalization to the real-world probability distribution $\mathcal{E}_{\text{real}} \subset \mathcal{E}_{\text{all}}$ is desired. This is called out-of-distribution (OOD) generalization. The true performance of a model can be defined as the expected risk under the true distribution of the data. The risk of a model $f$ under the environment $e$ with loss function $\mathcal{L}$ is defined as

$$R^e(f) = \mathbb{E}_{X^e, Y^e}[\mathcal{L}(f(X^e), Y^e)]. \qquad (2)$$

---

[*]Contact author: franck.rothen@unige.ch
[†]Contact author: samuel.klein@unige.ch
[‡]Contact author: matthew.leigh@unige.ch
[§]Contact author: tobias.golling@unige.ch

The out-of-distribution risk $R^{\text{OOD}}(f)$ of a model $f$ is defined as the maximum risk over all environments. In many cases where the dataset is large, it is common to approximate the out-of-distribution risk with the empirical risk, defined as

$$R^{\text{OOD}}(f) \approx R^D(f) = \max_{e \in \mathcal{E}_{\text{Tr}}} R^e(f). \qquad (3)$$

Discrepancies among different Monte Carlo simulators for the same process can serve as a proxy for assessing a model's generalization capabilities. If the individual environments diverge strongly, the empirical risk approximation is inadequate, and the model is likely to also perform poorly on real data.

Research on methods to improve generalization is ongoing [20–23]. However, this study focuses on a different approach, utilizing the geometric properties of the loss landscape. The generalization properties of a neural network are closely tied to the width of the local minimum in the loss landscape [24–26]. Specifically, sharp minima are typically found in regions where the loss varies significantly with small deviations. Such variations can be directly linked to model complexity and potential overfitting. According to the minimum-description-length principle in information theory, a model that requires fewer bits of information, and thus less complexity to describe the data, is likely to generalize better [27].

The empirical risk is assumed to closely approximate the true risk, albeit with minor distortions and shifts. The difference between the two, denoted as $\Delta R$, is statistically expected to be larger for a narrow minimum than for a wider one—i.e., $\Delta R_N \gtrsim \Delta R_W$. This is attributed to the increased sensitivity of the narrow minimum to minor deviations in the loss landscape. Consequently, the sharpness of the minimum can serve as an indicator of a model's generalization performance.

## III. SHARPNESS DEFINITION AND ADVERSARIAL ATTACKS

A local minimum $b$ is sharper than a local minimum $a$ if, for every perturbation strength $\epsilon$, the expected loss increase for $b$ is higher than for $a$:

$$\mathbb{E}_{\|\delta\|=\epsilon}[\Delta \mathcal{L}_a(\delta)] \leq \mathbb{E}_{\|\delta\|=\epsilon}[\Delta \mathcal{L}_b(\delta)], \quad \forall\, \epsilon \in \mathbb{R}_+, \qquad (4)$$

where $\Delta \mathcal{L}_i(\delta) \coloneqq \mathcal{L}_i(x + \delta) - \mathcal{L}_i(x)$ is the loss increase due to the perturbation $\delta$ for the local minimum $i$.

Bayesian approaches [28–30] and Lipschitz models [31,32] have been shown to reduce sharpness. However, the most straightforward way to exploit this characteristic is through adversarial white-box attacks.

A perfect adversarial attack involves modifying the input with a perturbation $\delta$ of bounded magnitude $\|\delta\| < \epsilon$, in order to maximize the corresponding loss,

$$\max_{\|\delta\|<\epsilon} \mathcal{L}(w, x + \delta, y), \qquad (5)$$

where $w$ is the model's weight, $x$ is the input, and $y$ is the target label. Adversarial robustness and low sharpness can be achieved by training the model directly on adversarial samples. This is known as adversarial training [33]. Evidently, increase in robustness comes at the cost of a decrease in natural accuracy [34]. The goal is to minimize the loss for the worst possible norm-constrained perturbation $\delta$.

It is straightforward to show that adversarial robustness is also a form of noise robustness. We assume the expected noise magnitude to be bounded by $\epsilon$. If $\delta$ is the $\epsilon$-bounded perturbation, which maximizes the model's prediction error, then by definition,

$$\forall\, \delta' \in \mathbf{B}_\epsilon, \quad \mathcal{L}(w, x + \delta', y) \leq \mathcal{L}(w, x + \delta, y), \qquad (6)$$

where $\mathbf{B}_\epsilon$ is the ball of radius $\epsilon$ centered around zero. Thus, the training loss for every possible noise possibility is necessarily smaller than the equivalent loss of an attacked parameter. Consequently, since the latter loss is being minimized by adversarial training, the model prediction also becomes significantly better in the case of noisy inputs. However, solving for the optimal perturbation $\delta$ is not computationally feasible for a high number of dimensions. Approximations of this perturbation $\delta$ are required. In the context of this paper, four different first-order adversarial training methods are considered.

The FGSM [33,35] attack is the most common adversarial attack. A fast optimal max-norm-constrained perturbation $\delta_{\text{FGSM}}$ can be obtained by the first-order Taylor expansion. In this approach, samples are transformed into their adversarial equivalents using

$$x \to x' = x + \epsilon \cdot \text{sign}(\nabla_x \mathcal{L}(w, x, y)). \qquad (7)$$

This transformation requires knowledge of the loss function gradient with respect to the inputs $x$. In a white-box scenario, where the model weights and architecture are known, this information can be easily obtained using traditional backpropagation. However, this comes with an increase in training time, as gradients must be calculated twice.

Alternatively, a superior approximation of the optimal norm-bound perturbation $\delta$ can be obtained by iterating the FGSM principle with smaller steps [36]. This iterative process is analogous to gradient descent. Just as the loss can be minimized, the loss increase caused by a perturbation $\delta$ can be maximized by inverse gradient descent, or gradient ascent. To prevent the perturbation from leaving the $\epsilon$-ball, a projection operator is applied after each step. This technique is known as projected gradient descent (PGD). A conceptual implementation for both FGSM and PGD is shown in Appendix B 1.

For a large iteration number $N$ and a small perturbation step $\alpha$, the PGD attack is expected to converge toward a local maximum within the $\epsilon$-ball around $x$. However, this is not necessarily the global maximum of the $\epsilon$-ball, which is a fundamental limitation of a first-order adversary. PGD significantly surpasses FGSM in terms of adversarial robustness [36]. However, while increasing $N$ and decreasing $\alpha$ improves the quality of adversarial samples, it also substantially increases the computational cost, because the gradient must be recalculated at each step. Therefore, the selection of $N$ should balance computational cost and adversarial robustness.

Sharpness-aware minimization (SAM) is an alternative method designed to increase the robustness of a model [37–39]. Similarly to adversarial training, this robustness is also obtained by applying a norm-constrained perturbation $\epsilon$, but on the model's weights $w$ instead of the data samples. In order to respect the original paper's notation, a perturbation in the weight space is denoted by $\epsilon$ and bounded by the rho-ball, $\|\epsilon\| < \rho$.

Following the same reasoning as before, the optimal perturbation $\epsilon$ cannot reasonably be computed and is thus replaced by its Taylor first-order approximation,

$$\epsilon_{\text{SAM}}(w) = \rho \cdot \text{sign}(\nabla_w \mathcal{L}(w)). \qquad (8)$$

The associated effective gradient is then approximated as follows:

$$\nabla_w \mathcal{L}_{\text{SAM}}(w) \approx \nabla_w \mathcal{L}(w)|_{w+\epsilon(w)}. \qquad (9)$$

A possible implementation of SAM is given in Appendix B 2.

A less disruptive approach is given by the dynamic sparse sharpness-aware minimization (SSAM-D) method [40]. This technique was inspired by another study [24], which revealed that only about 5% of the parameter space exhibited sharp minima behavior. The remainder of the parameter space naturally converges to flat local minima during standard stochastic gradient descent (SGD). The goal of this method is to apply the SAM perturbation only to the parameters that are most likely to benefit from it. The SSAM loss is thus defined as

$$\mathcal{L}_{\text{SSAM}} \coloneqq \max_{\|\epsilon\|<\rho} \mathcal{L}(w + \epsilon \odot \mathbf{m}_w), \qquad (10)$$

where $\mathbf{m}_w$ is a binary mask, which is dynamically updated during training.

By perturbing only a subset of parameters, the training process becomes smoother, resulting in improved natural accuracy at the cost of minimal decrease in robustness. Moreover, certain hardware setups may benefit from this approach by eliminating computations on parameters that are unlikely to gain from the perturbation. While standard training requires one complete backward pass and the SAM training strategy requires two, the SSAM-D training strategy only needs one complete backward pass and a partial one.

The corresponding pseudocode and mask generation can be found in Appendix B 3.

Second-order approximation of the maximal loss defined by Eq. (5), such as Hessian-aware adversarial training, could be considered to obtain a better approximation of the ideal perturbation and avoid high-dimensional saddle points. However, due to the heavy cost of Hessian computation, second-order attacks are not considered in this paper and are left as future work.

## IV. LOCAL LOSS SHARPNESS EVALUATION

### A. General considerations

A method to assess the effectiveness of the different training strategies in reducing the sharpness of the local minimum is required [26]. For simplicity in notation but without loss of generality, only the weight space is considered in the following theoretical considerations. Since very high-dimensional spaces are considered, direct visualization of the entire landscape is not possible. Sharpness can be intuitively compared by quantifying the loss at different perturbation strengths $\rho$. The sharper the local minimum, the higher the loss increase for a given perturbation strength. The expectation for a specific $\rho$ value is given by averaging the loss to every norm-constrained perturbation combination,

$$\mathbb{E}[\mathcal{L}_\rho] = \frac{1}{S_\rho} \oint_{\|\epsilon\|=\rho} \mathcal{L}(w + \boldsymbol{\epsilon}, x, y) d^n\boldsymbol{\epsilon} \qquad (11)$$

$$\stackrel{\text{M.C}}{\approx} \frac{1}{N} \sum_{i=1}^{N} \mathcal{L}(w + \boldsymbol{\epsilon}_i, x, y), \qquad (12)$$

where $n = \dim(w)$ is the dimension of the weight space and $S_\rho$ is the normalization constant given by the $\rho$-ball surface. However, neither this integral nor its Monte Carlo approximation is computationally feasible for high-dimensional spaces. Therefore, two different methods are proposed in this study to approximate the sharpness.

### B. Gradient ascent path tracing

The first proposed method for visualizing sharpness draws inspiration from previous considerations and the theoretical adversarial loss given by Eq. (5). Instead of computing the average loss increase for different perturbation strength values, only the upper bound is considered as a proxy for the sharpness:

$$\max_{\epsilon \leq \rho} \mathcal{L}^a(w + \epsilon) \leq \max_{\epsilon \leq \rho} \mathcal{L}^b(w + \epsilon) \stackrel{\Rightarrow}{\approx} \mathbb{E}[\mathcal{L}^a_\rho] \leq \mathbb{E}[\mathcal{L}^b_\rho]. \quad (13)$$

This method significantly reduces the number of computations required, as the test set needs to be evaluated only once. The remaining task is to compute the maximum loss in the $\rho$-ball, as given by Eq. (13). This is equivalent to the adversarial problem and can be treated similarly. The projected gradient ascent is used to find the best approximation of the maximum loss in the $\rho$-ball. The loss increase can be analyzed by plotting its value $\mathcal{L}_i$ at each step $i$ of the gradient ascent. In order to facilitate visual comparison between curves, the loss is shifted to zero at the local minimum—i.e., $\Delta \mathcal{L}_i = \mathcal{L}_i - \mathcal{L}_0$.

### C. Hessian analysis

An alternative method involves directly evaluating the Hessian matrix of the loss function at the local minimum. An approximation of the loss function for small perturbations can be obtained through a Taylor expansion,

$$\mathcal{L}(w + \epsilon) = \mathcal{L}(w) + \nabla \mathcal{L}(w)^T \epsilon + \frac{1}{2} \epsilon^T H_\mathcal{L}(w) \epsilon + \mathcal{O}(\epsilon^3),$$

where $\nabla \mathcal{L}(w)$ is the gradient of $\mathcal{L}$ at $w$, and $H_\mathcal{L}(w)$ is the Hessian matrix of $\mathcal{L}$ at $w$. Evidently, the gradient $\nabla \mathcal{L}$ is expected to vanish as the model converges. Therefore, only the second order characterized by the Hessian matrix $H_\mathcal{L}$ remains. However, direct computation of $H_\mathcal{L}$ is unrealistic for large weight and feature spaces. Assumptions and technical implementation for computing an approximation of $H_\mathcal{L}$ are provided in Appendix A. In order to efficiently extract sharpness information from the Hessian, the eigenvalues are considered as a proxy.

The largest eigenvalues are obtained using the von Mises algorithm [41], as detailed in Appendix A. The resulting eigenvalues for the different learning strategies are directly compared with each other. The larger the maximal eigenvalue, the sharper the local minimum.

## V. EXPERIMENTAL SETUP

The classifiers presented in this study aim to automate the distinction between jets originating from specific signal and background processes captured by the ATLAS detector. A jet refers to a collimated spray of particles produced by the fragmentation and hadronization of a high-energy quark or gluon. The signal process is a Higgs boson decaying into a bottom-antibottom pair, $H \to b\bar{b}$, while the background processes are formed from gluon- or quark-initiated jets. The baseline architecture of all classifiers is described in Appendix C 1. The classifiers presented in this paper are all trained on the publicly available resimulation-based dataset (RS3L) [42,43].

The RS3L dataset is generated by reshowering simulated partons using different configurations and software. This property can be used to evaluate the generalization capabilities of models trained on different MC simulations of the same process. The different reshowering scenarios are listed in Table I.

Because of similarities between the first four scenarios (RS3L0 to RS3L3), only the nominal Pythia8 [44] simulation and Herwig7 [45,46] datasets are considered. The number of physical constituents in a jet varies from event to event, as shown in the corresponding distribution in Fig. 1.

A significant increase in the number of constituents can be observed for the Herwig dataset compared to Pythia. This is due to different approximation in the parton showering process and expected final-state radiation. The distribution of jet mass is particularly important. The associated distribution can be observed for the Pythia and Herwig training datasets in Fig. 2.

The Higgs resonance is slightly less sharp in the Herwig dataset compared to the Pythia dataset. The background QCD distribution in the Herwig dataset is broader and peaks at higher mass values compared to the Pythia dataset.

All adversarial hyperparameters are determined via a grid search and are detailed in Appendix C 2. Although the goal is to improve performance on cross-evaluation datasets, the hyperparameter search is conducted using only the respective training dataset to emulate real-world conditions where the target set is unknown.

Only features available in jet reconstruction from real measurements are used as input parameters. The final

TABLE I. Augmentation found in the resimulation-based dataset.

| Augmentation | Description |
| --- | --- |
| RS3L0 | Jet showered with Pythia8 (Nominal) |
| RS3L1 | Numerical seed alteration |
| RS3L2 | FSR probability scale change by $1/\sqrt{2}$ |
| RS3L3 | FSR probability scale change by $\sqrt{2}$ |
| RS3L4 | Using Herwig7 as parton shower |

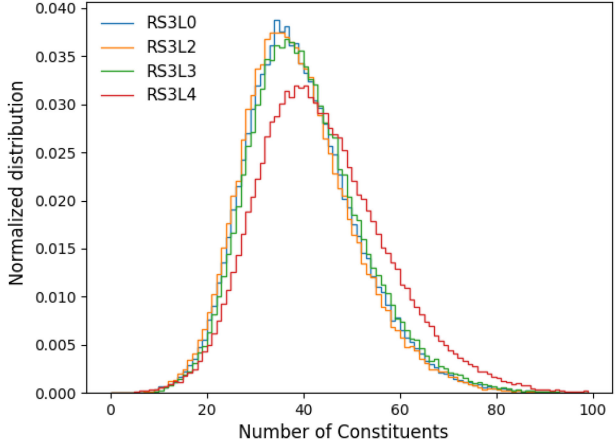

FIG. 1. Distribution of the number of constituents in a jet for the Pythia (RS3L0) and Herwig (RS3L4) training datasets.

variables used during training and their respective descriptions are listed in Table II.

The features can be segmented into two distinct groups. The first group, referred to as high-level features, represents the jet as a whole, such as the reconstructed mass or energy and its transverse momentum. The second group comprises low-level features, which describe individual constituents composing the jet.

In order to ensure a fair practical comparison between the different training methods, all models are trained on the same hardware, and a fixed time limit of 24 h is imposed.

All the code used to run these experiments is publicly available at [47].

## VI. RESULTS

### A. Cross-evaluation between Monte Carlo simulations

In order to evaluate the generalization capabilities of models trained under normal conditions, a cross-evaluation between the different Monte Carlo simulations, Pythia and Herwig, is performed. In addition to the area under the curve (AUC), the numerical background rejection, defined as the inverse of the false positive rate at a fixed signal efficiency $\epsilon_S$, is used as a performance metric. The resulting scores of the cross-validation for $\epsilon_S = 0.85$ are presented in Table III. These values are obtained by the seed average and standard deviation over five runs with different random seeds.

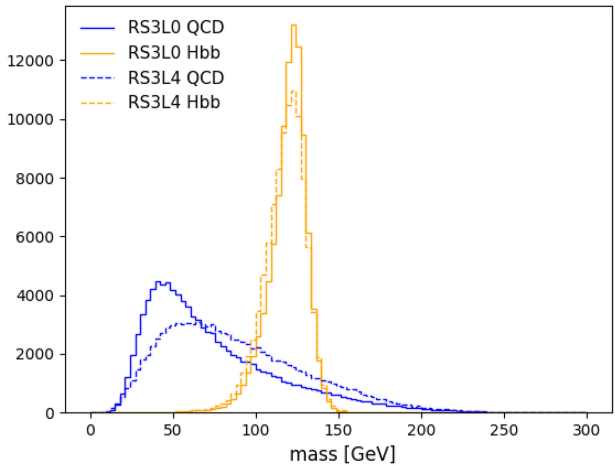

FIG. 2. Distribution of the jet mass for the Pythia (RS3L0) and Herwig (RS3L4) training datasets.

TABLE II. Model input features for the jet and particle constituents.

| Jet features | |
|---|---|
| Feature | Description |
| Log $p_T$ | Logarithm of the jet transverse momentum |
| Log $m$ | Logarithm of the jet mass |

| Particle constituent features | |
|---|---|
| Feature | Description |
| Log $p_T$ | Logarithm of the transverse momentum |
| Log $(p_T/p_{\text{Jet}})$ | Logarithm of $p_T$ normalized with respect to the jet |
| Log $E$ | Logarithm of the energy |
| Log $(E/E_{\text{Jet}})$ | Logarithm of $E$ normalized with respect to the jet |
| $\Delta\eta$ | Pseudorapidity difference relative to the jet |
| $\Delta\phi$ | Azimuthal angle difference relative to the jet |
| $\Delta R$ | Distance from the jet axis in the $\eta$-$\phi$ plane |
| Charge | Charge of the particle |
| Tanh $d_0$ | Transverse impact parameter (tanh) |
| Tanh $dz$ | Longitudinal impact parameter (tanh) |
| IsPhoton | Photon binary indicator |
| IsMuon | Muon binary indicator |
| IsElectron | Electron binary indicator |
| IsCH | Charged hadron binary indicator |
| IsNH | Neutral hadron binary indicator |

As expected, the results show that models trained and evaluated on different datasets perform significantly worse than those trained and evaluated on the same dataset. Given that both Pythia and Herwig simulate the same process, this discrepancy likely stems from overfitting to simulation. This suggests that the model may not effectively generalize to real-world data, underscoring the need for methods to enhance the model's generalization properties.

TABLE III. AUC and rejection at $\epsilon_S = 0.85$ signal efficiency for the different training and evaluation scenarios of the default model. Bold values indicate the highest performance for each metric.

| | Evaluation sets | | | |
|---|---|---|---|---|
| | Pythia | | Herwig | |
| Training sets | AUC | Rejection | AUC | Rejection |
| Pythia | **0.978 ± 0.002** | **24.2 ± 0.4** | 0.953 ± 0.004 | 11.3 ± 0.2 |
| Herwig | 0.965 ± 0.003 | 15.0 ± 0.2 | **0.974 ± 0.002** | **21.2 ± 0.4** |

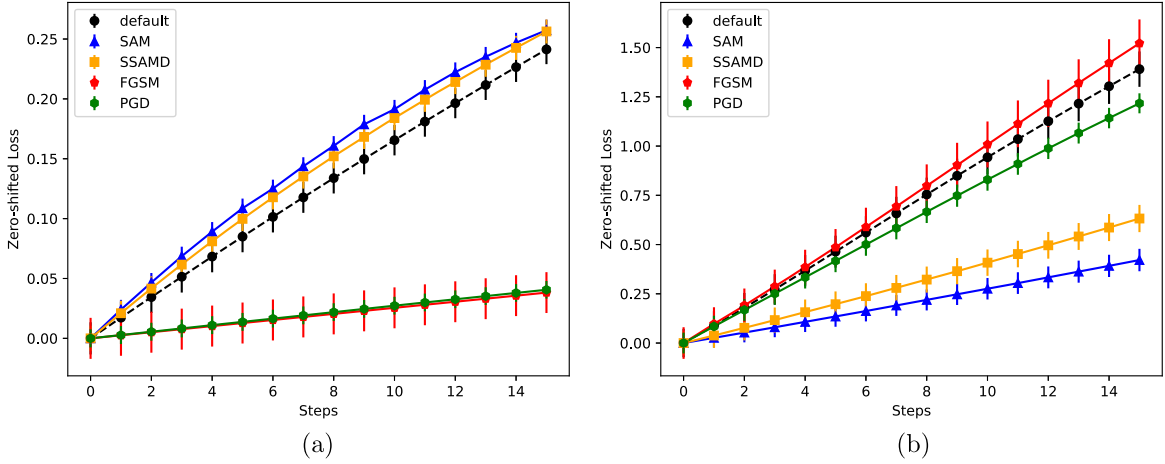

FIG. 3. Comparison of gradient ascent on the Pythia dataset for default and adversarial trained dense networks. (a) Feature-space. (b) Weight-space.

### B. Analysis of adversarial sharpness reduction

The same models are trained, but using the adversarial training strategies described in Sec. III. The sharpness of the local minima is analyzed following the methods presented in Sec. IV. The resulting plots of the gradient ascent for models trained and evaluated on the nominal Pythia dataset are shown in Fig. 3. Similar plots for all cross-evaluation scenarios are shown in Appendix D 3 in Fig. 7. The values and uncertainties are given by the seed average and associated standard deviation.

As desired, training using adversarial samples generated by either FGSM or PDG results in a significantly flatter local minimum in feature space compared to default training. The same behavior is also observed in the weight space for both sharpness-aware minimization methods. It is interesting to note that feature space and weight space sharpness are visibly not correlated. While FGSM and PDG both reduce the sharpness in the feature space, SAM and SSAMD do not. The inverse is true for the weight space.

The results of the Hessian eigenvalue analysis for models trained and evaluated on Pythia are shown in Table IV. The equivalent for all cross-evaluation scenarios is shown in Appendix D 3 in Table X.

TABLE IV. Largest Hessian eigenvalues for the different training methods and perturbation spaces for models trained and evaluated on Pythia. Lower values correlate with wider minimas. Bold values indicate the smallest eigenvalues (widest minimas).

| Methods | Feature space | | Weight space | |
| --- | --- | --- | --- | --- |
| | Hbb | QCD | Hbb | QCD |
| Default | $0.84 \pm 0.08$ | $0.03 \pm 0.01$ | $0.31 \pm 0.05$ | $0.28 \pm 0.07$ |
| SAM | $0.82 \pm 0.11$ | $0.07 \pm 0.04$ | $\mathbf{0.11 \pm 0.01}$ | $\mathbf{0.12 \pm 0.01}$ |
| SSAMD | $0.98 \pm 0.09$ | $0.04 \pm 0.01$ | $0.22 \pm 0.01$ | $0.19 \pm 0.03$ |
| FGSM | $0.17 \pm 0.01$ | $0.024 \pm 0.006$ | $0.80 \pm 0.09$ | $0.49 \pm 0.07$ |
| PGD | $\mathbf{0.056 \pm 0.004}$ | $\mathbf{0.005 \pm 0.002}$ | $0.72 \pm 0.07$ | $0.42 \pm 0.08$ |

Compared to the default training method, the largest eigenvalue of the classification layer Hessian is significantly reduced when using the SAM and SSAM-D training strategies. Conversely, the FGSM and PGD methods result in a significant decrease in the largest eigenvalue of the feature-space Hessian. Additionally, the PGD method expectedly results in a stronger decrease in sharpness than the FGSM method, as it generates more accurate adversarial samples.

### C. Comparison of generalization performance

The previous section established that the implemented methods successfully result in wider minima. The corresponding generalization performance can be evaluated once again by cross-evaluation on both Pythia and Herwig datasets. To quantify a model's performance increase compared to the default training strategy, the fractional relative performance increase $\Delta S$ is introduced. This metric is defined as

$$\Delta S = \frac{S_a'^b - S_a^b}{S_b^b - S_a^b}, \qquad (14)$$

where $S_i^j$ is the score of the default model trained on dataset $i$ and evaluated on dataset $j$, and $S'$ is the score of the considered method. It is trivial to verify that $\Delta S$ becomes 0 if the model performs as well as the default model, and 1 if the model performs as well as the theoretical best model. More generally, any strictly positive $\Delta S$ indicates an improvement over the default model.

This metric is only valid under the condition that the expected score difference between the default score and the optimal score is significantly larger than the statistical fluctuations surrounding these scores. As was given by the reference results in Table III, this condition is largely satisfied. Any score metric for $S$ can be used, but AUC is preferred, since it is a global performance metric. The resulting performance increase for the different adversarial training methods is shown in Table V. The corresponding raw results and associated rejection curves for the entire signal efficiency $\epsilon_S$ spectrum can be found in Appendix D 1, specifically in Table IX and Fig. 4, respectively.

TABLE V. Fractional generalization performance increase $\Delta S$ for the different adversarial training methods for both the AUC and rejection at the $\epsilon = 0.85$ metric. (Training set $\rightarrow$ Evaluation set). Bold values indicate the largest performance improvements.

|  | Pythia $\rightarrow$ Herwig | | Herwig $\rightarrow$ Pythia | |
| --- | --- | --- | --- | --- |
|  | $\Delta_{AUC}$ | $\Delta_{Rej}$ | $\Delta_{AUC}$ | $\Delta_{Rej}$ |
| Default | 0 | 0 | 0 | 0 |
| SAM | $0.44 \pm 0.02$ | $0.25 \pm 0.02$ | $0.20 \pm 0.01$ | $0.07 \pm 0.01$ |
| SSAMD | **$0.47 \pm 0.01$** | **$0.29 \pm 0.02$** | $0.23 \pm 0.01$ | $0.14 \pm 0.01$ |
| FGSM | $0.21 \pm 0.01$ | $0.03 \pm 0.01$ | $0.44 \pm 0.02$ | $0.27 \pm 0.02$ |
| PGD | $0.46 \pm 0.02$ | $0.25 \pm 0.02$ | **$0.76 \pm 0.03$** | **$0.62 \pm 0.02$** |
| Target | 1 | 1 | 1 | 1 |

The results show that all adversarial training methods result in a significant improvement in generalization performance, supporting the hypothesis of flatter minimum having better generalization properties. As anticipated, PGD performs better than FGSM in all scenarios. This is because samples generated using PGD attacks are generally a better approximation to perfect adversarial samples, as given by definition in Eq. (5). A larger computational cost is, however, required.

It is observed that feature-space attack methods generalize better from Herwig to Pythia than from Pythia to Herwig, whereas weight-space attacks show the opposite trend. This behavior can be partially attributed to mass correlation, as discussed in Appendix D 2.

The rejection performance as a function of the signal efficiency $\epsilon_S$ for models trained and evaluated on identical simulators is discussed in Appendix D 1.

## VII. CONCLUSION

This study suggests that divergences may emerge when models are trained on two distinct MC simulations of the same processes. Simple models achieve good performance on their respective training sets, but cross-evaluation scenarios reveal a lack of generalization capabilities. This discrepancy likely arises from overfitting to simulation artifacts, indicating that supervised training does not capture the underlying physical processes.

Generalization performance is closely related to the shape of the loss landscape around the considered local minimum. Adversarial methods were used to exploit this property, and their effectiveness in decreasing loss sharpness was demonstrated. All implemented adversarial strategies resulted in a significant increase in generalization performance. The projected gradient descent method was found to be the overall most effective generalization method.

The results of this study are promising, and they also highlight areas for future exploration and improvement. The most crucial future work involves further verification of how well models trained on Monte Carlo simulations can generalize to real-world data. Depending on how significant the generalization issue is, generalization strategies may significantly enhance the power of discriminators in HEP.

## ACKNOWLEDGMENTS

The authors would like to acknowledge funding through SNSF Sinergia Grant No. CRSII5_193716 "Robust Deep Density Models for High-Energy Particle Physics and Solar Flare Analysis (RODEM)" and SNSF Project Grant No. 200020_212127 "At the two upgrade frontiers: machine learning and the ITk Pixel detector."

## DATA AVAILABILITY

The data that support the findings of this article are openly available [42].

## APPENDIX A: HESSIAN APPROXIMATION

Consider $\mathcal{L}: \mathbb{R}^n \rightarrow \mathbb{R}$ and recall the definition of the Hessian, which is the matrix of second-order partial derivatives of the loss function $\mathcal{L}$,

$$(H_{\mathcal{L}})_{ij} = \frac{\partial^2 \mathcal{L}}{\partial w_i \partial w_j} = \frac{\partial}{\partial w_i}\left(\frac{\partial \mathcal{L}}{\partial w_j}\right). \quad \text{(A1)}$$

The right-hand term, $\partial_{w_j}\mathcal{L}$, can easily be obtained through backpropagation. Therefore, the entire Hessian matrix, $(H_{\mathcal{L}})_{ij} = \partial_{w_i}(\partial_{w_j}\mathcal{L})$, can be obtained by $n$ additional backpropagation passes, to calculate all the $n^2$ second-order partial derivatives.

However, the memory and computational cost of these calculations is prohibitive for models with very large weight spaces. Therefore, to reduce the computational cost considerably, only subspaces of the feature and weight space can be considered. In the case of the feature space, only the $n_C = 5$ first constituents out of the $N_C = 40$ total constituents are considered. This approximation is reasonable, as the leading constituents contribute the most to the loss function. Since robustness is significantly impacted by the last classification layer [48], only this layer is considered in the case of the weight space. While not considering the entire space, the reduced Hessian still provides a good approximation of the overall sharpness of the local minimum.

Furthermore, since an infinitely differentiable loss function is considered, computation of the Hessian matrix can be further simplified by considering Schwarz's theorem. This theorem states that the order of differentiation does not matter for smooth $C^2$ functions:

$$\frac{\partial^2 \mathcal{L}}{\partial w_i \partial w_j} = \frac{\partial^2 \mathcal{L}}{\partial w_j \partial w_i}. \tag{A2}$$

Therefore, the Hessian matrix is symmetric, and only $n(n+1)/2$ second-order partial derivatives need to be calculated. This further reduces the computational cost.

Having established an effective Hessian matrix $H_\mathcal{L}$, the next step is to extract sharpness information. This can be achieved by considering the eigenvalues, as they represent local curvature in the associated eigendirection. Nonsymmetric eigenvalue algorithms such as the QR method could be implemented for a full description, but they are too complicated for this use case. Instead, the focus is set on the computation of the largest eigenvalue. This value serves as an upper bound for the local sharpness. A lower maximal eigenvalue directly indicates reduced sharpness. To achieve the desired computation, the power iteration method, also referred to as the von Mises iteration algorithm, can be used:

$$b_k = \frac{H_\mathcal{L} b_{k-1}}{\|H_\mathcal{L} b_{k-1}\|} = \frac{H_\mathcal{L}^k b_0}{\|H_\mathcal{L}^k b_0\|}, \tag{A3}$$

where $b_0$ is a random vector and $b_k$ is the $k$th approximation of the eigenvector with the largest eigenvalue.

The associated eigenvalue $\lambda_k$ can easily be obtained by evaluating the Rayleigh quotient,

$$\lambda_k = \frac{b_k^T H_\mathcal{L} b_k}{b_k^T b_k}. \square \tag{A4}$$

The previously mentioned QR method is a generalization of this procedure.

## APPENDIX B: IMPLEMENTATION OF ADVERSARIAL TRAINING

### 1. Fast gradient method and projected gradient descent

A possible implementation of adversarial training using the FGSM and PGD method is found in Algorithm 1.

In this context, $N$ is the number of steps; $\Pi_{x,\epsilon}$ is the ball-projection operator, centered around $x$ with radius $\epsilon$; and $\alpha$ is the step size. This last parameter is comparable to a learning rate. To allow the perturbation to reach the maximum possible magnitude, $\alpha$ should be chosen such that $\alpha \cdot N \geq \epsilon$. Notably, the FGSM attack can be considered a special case of the PGD attack, where $N = 1$ and $\alpha = \epsilon$.

### 2. Sharpness-aware minimization (SAM)

Compared to the FGSM attack, an additional step is required for the SAM approach. While the optimal perturbation $\epsilon_{\text{SAM}}$ on the weight space is obtained through Eq. (8), it is not evident how to use it during a learning step. Following the original paper's [37] derivation, the gradient induced by the SAM loss function, $\nabla_w \mathcal{L}_{\text{SAM}}(w)$, is reconsidered. The value for $\epsilon_{\text{SAM}}$ can be inserted back into the loss term to obtain

$$\begin{aligned}\nabla_w \mathcal{L}_{\text{SAM}}(w) &= \nabla_w \mathcal{L}(w + \epsilon(w)) \\ &= \frac{d(w+\epsilon(w))}{dw} \nabla_w \mathcal{L}(w)|_{w+\epsilon(w)} \\ &= \nabla_w \mathcal{L}(w)|_{w+\epsilon(w)} + \frac{d\epsilon(w)}{dw} \mathcal{L}(w)|_{w+\epsilon(w)}. \end{aligned} \tag{B1}$$

For computational acceleration reasons, the expression is then further simplified by dropping the second-order term, thus obtaining

$$\nabla_w \mathcal{L}_{\mathcal{SAM}}(w) \approx \nabla_w \mathcal{L}(w)|_{w+\epsilon(w)}. \tag{B2}$$

An implementation of SAM would therefore look like Algorithm 2.

**Algorithm 1.** Projected gradient descent (PGD).

---

**function** Train using PGD (weights **w**, training set $\mathcal{S}$, perturbation strength $\epsilon$, learning rate $\eta$, number of epoch $T$, number of PGD steps $N$, step size $\alpha$)

    **for** epoch $t = 1, 2, \ldots T$ **do**
        **for** batches $\mathcal{B} \in \mathcal{S}$ **do**
            $\mathbf{x}' \leftarrow \mathbf{x}$
            **for** PGD steps $n = 1, 2, \ldots N$ **do**
                # Compute $\delta$ via Eq. (7)
                $\delta \leftarrow \alpha \cdot \text{sign}(\nabla_\mathbf{x} \mathcal{L}(\mathbf{w}, \mathbf{x}, \mathbf{y}))$
                # Apply perturbation
                $\mathbf{x}' \leftarrow \mathbf{x}' + \delta$
                # Project back to $\epsilon$-ball
                $\mathbf{x}' \leftarrow \Pi_{\mathbf{x},\epsilon}(\mathbf{x}')$
            **end for**
            # Update weights
            $\mathbf{w} \leftarrow \mathbf{w} - \eta \nabla_\mathbf{w} \mathcal{L}(\mathbf{w}, \mathbf{x}', \mathbf{y})$
        **end for**
    **end for**
**end function**

**Algorithm 2.** Sharpness-aware minimization (SAM).

---

**function** Train USING SAM (weights **w**, training set $\mathcal{S}$, perturbation strength $\rho$, learning rate $\eta$, number of epoch $T$)

    **for** epoch t $= 1, 2, \ldots T$ **do**
        **for** batches $\mathcal{B} \in \mathcal{S}$ **do**
            # Compute perturbation $\epsilon$ via Eq. (8)
            $\epsilon \leftarrow \rho \cdot \text{sign}(\nabla_\mathbf{w} \mathcal{L}(\mathbf{w}))$
            # Update weights
            $\mathbf{w} \leftarrow \mathbf{w} - \eta \nabla_\mathbf{w} \mathcal{L}(\mathbf{w})|_{\mathbf{w}+\epsilon(\mathbf{w})}$
        **end for**
    **end for**
**end function**

### 3. Dynamic sparse sharpness-aware minimization (SSAM-D)

Dynamic sparse sharpness-aware minimization aims to reduce the disruption caused by the adversarial training by applying an adversarial attack only on a masked selection of weights. The challenge lies in determining an optimal mask. Fisher information and dynamic sparse masking are two common options for generating the mask [40]. Due to their similarity, only the second option is implemented in this study. This approach is characterized by a perturbation-mask-dropping phase, followed by a perturbation-mask-growth phase. At regular intervals, the mask is updated to remove the $N_{\text{drop}}$ flattest parameters,

$$\mathbf{m}'_{k+1} = \mathbf{m}_k - \underset{w \in \mathbf{m}_k}{\text{ArgBottom}}(|\nabla \mathcal{L}(w)|, N_{\text{drop}}), \quad \text{(B3)}$$

where the $\text{Bottom}_{a \in \mathcal{S}}(f(a), N)$ function returns the $N$ smallest values of $f(a)$ for $a \in \mathcal{S}$. This is followed by the growth phase, where a random selection of $N_{\text{grow}}$ parameters is added to the mask,

$$\mathbf{m}_{k+1} = \mathbf{m}'_{k+1} + \underset{w \notin \mathbf{m}_k}{\text{Random}}(N_{\text{growth}}), \quad \text{(B4)}$$

where the function $\text{Random}_{\mathcal{S}}(N)$ returns a random selection of $N$ elements from the set $\mathcal{S}$. To maintain the sparsity $S$ of the mask constant, the number of parameters dropped and added should be equal: $N_{\text{drop}} = N_{\text{grow}}$. A possible implementation of SSAM-D is given in Algorithm 3.

Algorithm 3. Dynamic sparse SAM.

---

**function** Train USING SSAM-D (weights $\mathbf{w}$, sparse ratio $S$, number of epoch $T$, update interval $T_m$, learning rate $\eta$, training set $\mathcal{S}$, perturbation strength $\rho$, number of parameters to drop $N_{\text{drop}}$, number of parameters to grow $N_{\text{grow}}$)

---

    **for** epoch t = 1, 2, …T **do**
        **for** batches $\mathcal{B} \in \mathcal{S}$ **do**
            # Compute $\epsilon$ via Eq. (8)
            $\epsilon \leftarrow \rho \cdot \text{sign}(\nabla_{\mathbf{w}} \mathcal{L}(\mathbf{w}))$
            **if** $t \bmod T_m = 0$ **then**
                # Regenerate $\mathbf{m}$ via Eqs. (B3) and (B4)
                $\mathbf{m}_d \leftarrow \underset{w \in \mathbf{m}}{\text{Arg Bot}}(|\nabla \mathcal{L}(w)|, N_{\text{drop}})$
                $\mathbf{m} \leftarrow \mathbf{m} - \mathbf{m}_d$
                $\mathbf{m} \leftarrow \mathbf{m} + \underset{w \notin \mathbf{m}}{\text{Random}}(N_{\text{growth}})$
            **end if**
            # Apply mask
            $\epsilon \leftarrow \epsilon \odot \mathbf{m}$
        **end for**
        # Update weights
        $\mathbf{w} \leftarrow \mathbf{w} - \eta \nabla_{\mathbf{w}} \mathcal{L}(\mathbf{w} + \epsilon)$
    **end for**
**end function**

TABLE VI. Simplified dense network architecture and number of parameters.

| Layer | Type | Parameters |
|---|---|---|
| Input | Input vector (602) | … |
| First hidden | Dense (128) | 77,184 |
|  | Batch normalization | 256 |
| Second hidden | Dense (64) | 8,256 |
|  | Batch normalization | 128 |
| Final hidden | Dense (32) | 2,080 |
|  | Batch normalization | 64 |
| Output | Dense (1) | 33 |
| Total |  | 88,001 |

## APPENDIX C: TECHNICAL IMPLEMENTATION

### 1. Model architecture

The final network was deliberately kept simple to ensure that complex behaviors do not interfere with the fair comparison of the training strategies. Therefore, a traditional dense model is chosen and consists of three hidden layers of sizes 128, 64, and 32, respectively. Given 40 constituents with fifteen features each and two jet features, the input layer has 602 neurons. The output layer has a single neuron, representing binary classification. An additional bias node is added to every major layer. The complete architecture for the dense network is summarized in Table VI.

### 2. Hyperparameters

In addition to the underlying model architecture given in Appendix C 1, the general hyperparameters used for all models are listed in Table VII. The adversarial hyperparameters for the different methods are listed in Table VIII.

## APPENDIX D: COMPLEMENTARY RESULTS

### 1. Raw performance score

The non-normalized performance score $S$ for the different adversarial training methods is shown in Table IX.

The results presented in Sec. VI C are reported at a fixed signal efficiency of $\epsilon_S = 0.85$. In order to verify if similar results are obtained across the entire signal efficiency spectrum, the rejection in function of signal efficiency is presented in Figs. 4 and 5.

TABLE VII. General hyperparameters.

| Parameter | Value |
|---|---|
| Number of jets per class | 500000 |
| Number of constituents | 40 |
| Batch size | 128 |
| Optimizer | AdamW |
| Learning rate | 0.0001 |

TABLE VIII. Adversarial hyperparameters.

| FGSM | | SAM | |
|---|---|---|---|
| Parameter | Value | Parameter | Value |
| Epsilon | 0.007 | Rho | 0.7 |
| PGD | | SSAMD | |
| Parameter | Value | Parameter | Value |
| Epsilon | 0.007 | Rho | 0.1 |
| Number of steps | 5 | Sparsity | 0.5 |
| Step size | 0.01 | Update frequency | 5 |
| | | Drop rate | 0.5 |

TABLE IX. Raw performance score $S$ for the different adversarial training methods for both AUC and rejection at $\epsilon = 0.85$ metric. (Training set → Evaluation set). Bold values indicate the highest raw performance scores.

| | Pythia → Herwig | | Herwig → Pythia | |
|---|---|---|---|---|
| | $S_{AUC}$ | $S_{Rej}$ | $S_{AUC}$ | $S_{Rej}$ |
| Default | $0.953 \pm 0.004$ | $11.3 \pm 0.2$ | $0.965 \pm 0.003$ | $15.0 \pm 0.1$ |
| SAM | $0.962 \pm 0.002$ | $13.8 \pm 0.1$ | $0.968 \pm 0.001$ | $15.6 \pm 0.2$ |
| SSAMD | $\mathbf{0.963 \pm 0.002}$ | $\mathbf{14.2 \pm 0.2}$ | $0.968 \pm 0.003$ | $16.3 \pm 0.2$ |
| FGSM | $0.957 \pm 0.004$ | $11.5 \pm 0.1$ | $0.971 \pm 0.002$ | $17.5 \pm 0.2$ |
| PGD | $0.962 \pm 0.003$ | $13.8 \pm 0.2$ | $\mathbf{0.975 \pm 0.003}$ | $\mathbf{20.7 \pm 0.3}$ |
| Target | $0.974 \pm 0.002$ | $21.2 \pm 0$ | $0.978 \pm 0.002$ | $24.2 \pm 0$ |

It can be observed that the ranking of the different training methods remains consistent across most of the signal efficiency spectrum, with PGD consistently outperforming other methods. Notably, models trained on Pythia and evaluated on Herwig show a significant performance boost at lower signal efficiencies compared to higher values.

The performances of the different training methods on the same simulators diverge only slightly from the default baseline training method. Sharpness-aware methods, especially SSAMD, results in a slightly improved performance average across the entire signal efficiency $\epsilon_S$ spectrum. The use of traditional FGSM attacks results in slightly worse average performance. These results are perfectly in line with theoretical predictions [34,49]. However, it can also be observed that PGD outperforms the default training method, indicating potential overfitting of the default model.

### 2. Mass correlation

An asymmetry is observed in Sec. VI C, where feature-space attack methods generalize better from Herwig to Pythia than from Pythia to Herwig, whereas weight-space attacks show the opposite trend. This behavior can be partially attributed to the impact of the different learning strategies on mass correlation. The mass sculpting is quantified using the Jensen-Shannon divergences and can be found in Fig. 6.

It can be observed that training with feature-space adversarial methods consistently results in higher mass correlation compared to weight-space adversarial methods, which do not deviate from the default training strategy. As shown in Fig. 2, the mass distributions of the Higgs signal and background diverge more for Pythia than for Herwig.

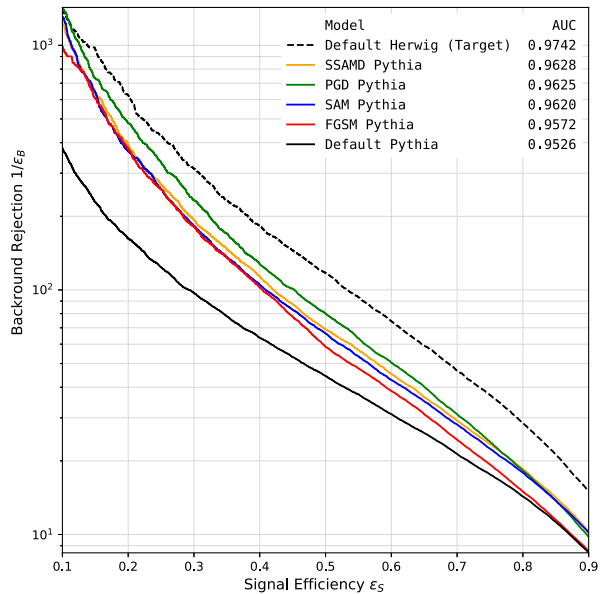
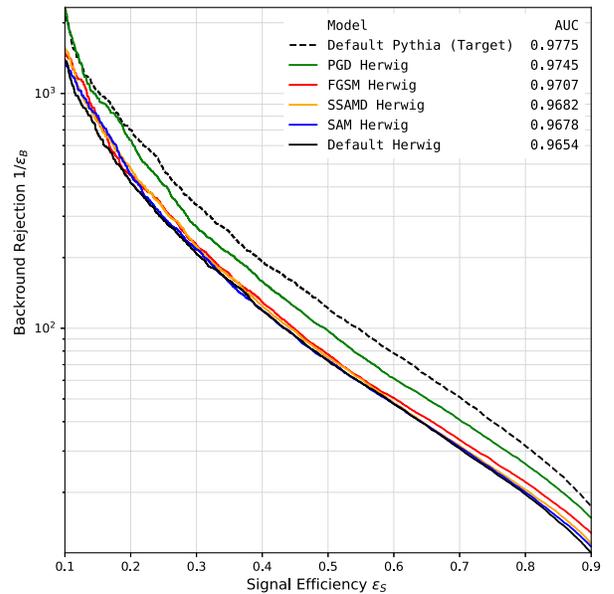

FIG. 4. Rejection in function of signal efficiency for the different training methods and cross-evaluation scenarios. (Training set → Evaluation set). (a) Pythia → Herwig. (b) Herwig → Pythia.

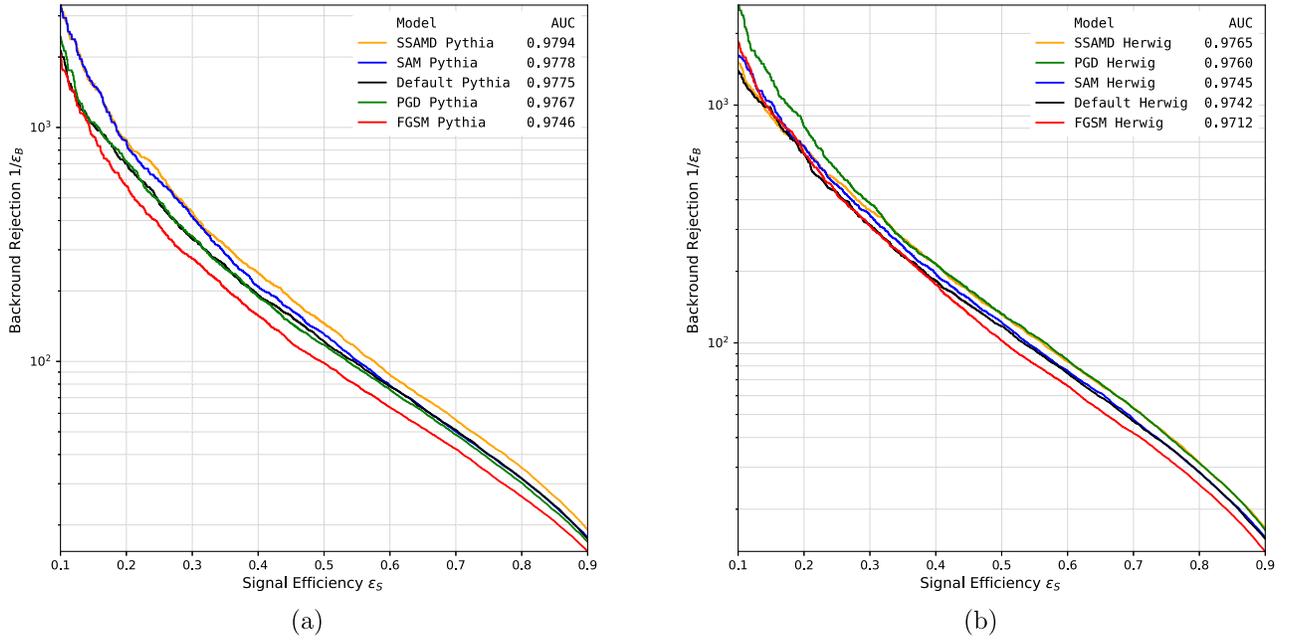

FIG. 5. Rejection in function of signal efficiency for the different methods trained and evaluated on the same simulator. (Training set → Evaluation set). (a) Pythia → Pythia. (b) Herwig → Herwig.

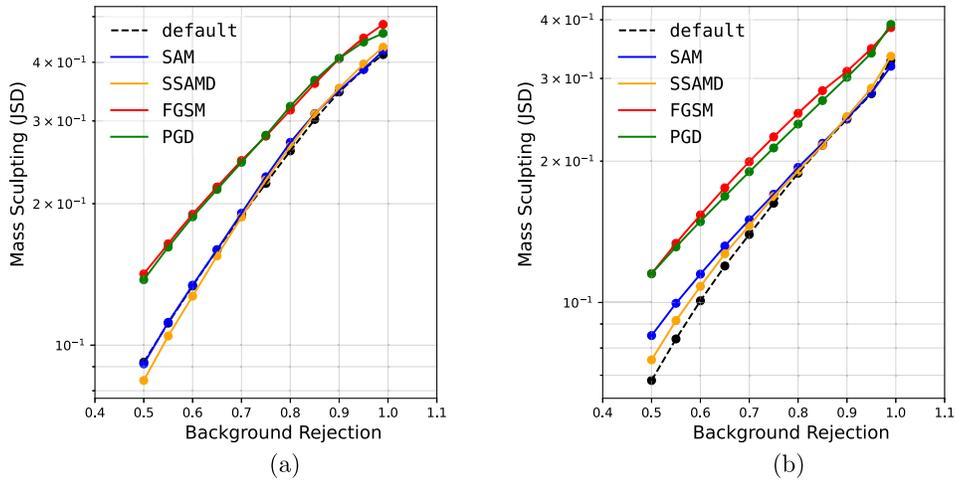

FIG. 6. Mass correlation (Jenson-Shannon divergence) between the signal and background rejection for the different training methods and cross-evaluation scenarios. (Training set → Evaluation set). (a) Pythia → Pythia. (b) Herwig → Herwig.

Therefore, the additional mass correlation naturally improves evaluation on Pythia compared to Herwig, explaining this asymmetry.

### 3. Sharpness reduction analysis

The sharpness reduction analysis, presented in Sec. VI B, only considers models trained on Pythia and evaluated on Pythia for brevity. However, in order to verify if the assumption of sharpness reduction holds, this needs to be verified for all cross-evaluation scenarios. The gradient ascent of all training strategies can be found for all cross-evaluation scenarios in Fig. 7. The largest Hessian eigenvalues for all methods and evaluation scenarios are summarized in Table X. It can be observed that both the gradient ascent and the Hessian analysis for the different training strategies are consistent across all cross-validation scenarios. Feature-space methods systematically reduce sharpness in feature space, while the equivalent is true for weight-space methods. Most notably, this remains true even in cases where the training set and evaluation set originate from different simulators, thus demonstrating the overall robustness of the training methods.

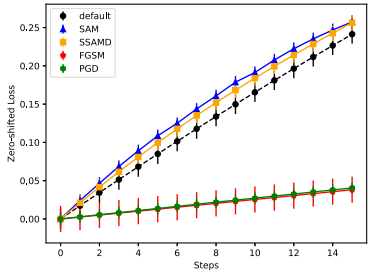
(Pythia → Pythia)

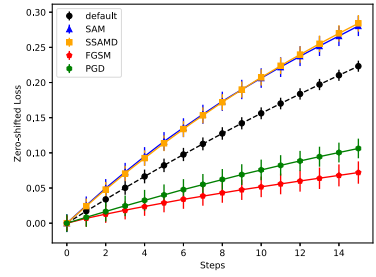
(Herwig → Herwig)

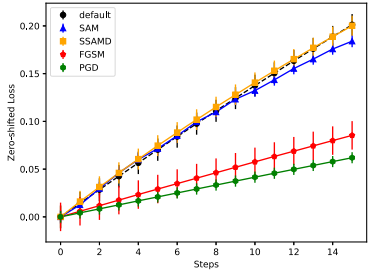
(Pythia → Herwig)

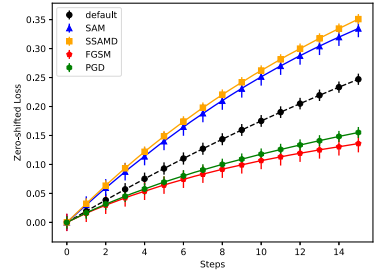
(Herwig → Pythia)

Feature-space

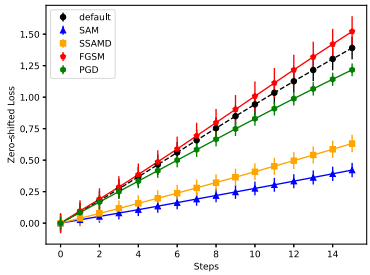
(Pythia → Pythia)

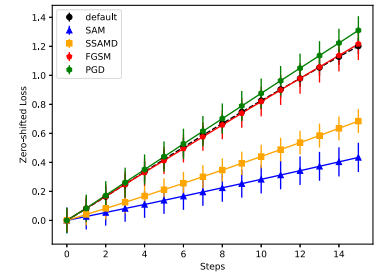
(Herwig → Herwig)

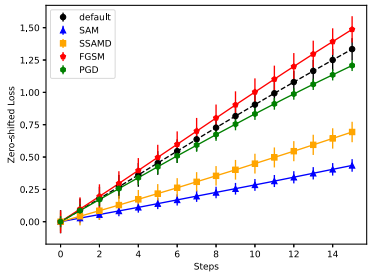
(Pythia → Herwig)

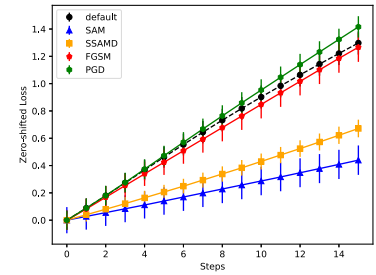
(Herwig → Pythia)

Weight-space

FIG. 7. Comparison of gradient ascent for default and adversarial trained dense networks for the different evaluation scenarios. (Training set → Evaluation set).

TABLE X. Largest Hessian eigenvalues for the different evaluation scenarios, training methods and perturbation spaces. Lower values correlate with wider minimas. (Training set → Evaluation set). Bold values indicate the smallest eigenvalues (widest minimas) for each scenario.

| Methods | Pythia → Pythia | | Herwig → Herwig | | Pythia → Herwig | | Herwig → Pythia | |
|---|---|---|---|---|---|---|---|---|
| | Hbb | QCD | Hbb | QCD | Hbb | QCD | Hbb | QCD |
| | | | | Feature space | | | | |
| Default | $0.84 \pm 0.08$ | $0.03 \pm 0.01$ | $0.95 \pm 0.16$ | $0.24 \pm 0.11$ | $0.58 \pm 0.06$ | $0.04 \pm 0.01$ | $1.23 \pm 0.41$ | $0.34 \pm 0.15$ |
| SAM | $0.82 \pm 0.11$ | $0.07 \pm 0.05$ | $0.95 \pm 0.26$ | $0.26 \pm 0.10$ | $0.57 \pm 0.08$ | $0.06 \pm 0.03$ | $0.78 \pm 0.33$ | $0.32 \pm 0.12$ |
| SSAMD | $0.98 \pm 0.09$ | $0.04 \pm 0.02$ | $0.86 \pm 0.15$ | $0.24 \pm 0.09$ | $0.64 \pm 0.07$ | $0.04 \pm 0.02$ | $0.75 \pm 0.21$ | $0.30 \pm 0.12$ |
| FGSM | $0.17 \pm 0.01$ | $0.02 \pm 0.01$ | $0.33 \pm 0.14$ | $0.11 \pm 0.05$ | $0.16 \pm 0.01$ | $0.04 \pm 0.01$ | $0.57 \pm 0.15$ | $0.09 \pm 0.05$ |
| PGD | $\mathbf{0.056 \pm 0.004}$ | $\mathbf{0.005 \pm 0.002}$ | $\mathbf{0.25 \pm 0.04}$ | $\mathbf{0.06 \pm 0.02}$ | $\mathbf{0.05 \pm 0.01}$ | $\mathbf{0.006 \pm 0.001}$ | $\mathbf{0.55 \pm 0.05}$ | $\mathbf{0.06 \pm 0.02}$ |
| | | | | Weight space | | | | |
| Default | $0.32 \pm 0.05$ | $0.28 \pm 0.07$ | $0.26 \pm 0.04$ | $0.26 \pm 0.05$ | $0.43 \pm 0.19$ | $0.56 \pm 0.20$ | $1.30 \pm 0.21$ | $0.23 \pm 0.07$ |
| SAM | $\mathbf{0.11 \pm 0.01}$ | $\mathbf{0.12 \pm 0.01}$ | $\mathbf{0.13 \pm 0.01}$ | $\mathbf{0.14 \pm 0.01}$ | $\mathbf{0.13 \pm 0.03}$ | $\mathbf{0.15 \pm 0.02}$ | $\mathbf{0.14 \pm 0.01}$ | $\mathbf{0.10 \pm 0.01}$ |
| SSAMD | $0.22 \pm 0.01$ | $0.20 \pm 0.03$ | $0.19 \pm 0.01$ | $0.18 \pm 0.03$ | $0.29 \pm 0.06$ | $0.28 \pm 0.07$ | $0.36 \pm 0.05$ | $0.13 \pm 0.03$ |
| FGSM | $0.80 \pm 0.09$ | $0.49 \pm 0.07$ | $0.72 \pm 0.11$ | $0.59 \pm 0.12$ | $1.01 \pm 0.11$ | $0.66 \pm 0.08$ | $2.15 \pm 0.27$ | $0.53 \pm 0.13$ |
| PGD | $0.72 \pm 0.07$ | $0.42 \pm 0.08$ | $0.57 \pm 0.07$ | $0.48 \pm 0.08$ | $0.92 \pm 0.09$ | $0.69 \pm 0.12$ | $1.91 \pm 0.26$ | $0.44 \pm 0.12$ |